\documentclass[10pt,aps,prl,reprint,groupedaddress,showkeys,superscriptaddress,amsmath,amssymb,onecolumn]{revtex4-2}

\pdfoutput=1

\usepackage[utf8]{inputenc}
\usepackage{amssymb,amsmath}
\usepackage{graphicx}
\usepackage{booktabs} 
\usepackage{hyperref}
\usepackage{csquotes} 
\usepackage{commath}
\usepackage{url}
\usepackage[american]{babel}

\newcommand{\nexp}{n_{\mathrm{E}}^{(\mathrm{exp})}}
\newcommand{\nlog}{n_{\mathrm{E}}^{(\mathrm{log})}}
\newcommand{\nbass}{n_{\mathrm{E}}^{(\mathrm{bass})}}
\newcommand{\ngeneral}{n_{\mathrm{E}}}
\newcommand{\thalfexp}{t_{1/2}^{(\mathrm{exp})}}
\newcommand{\thalflog}{t_{1/2}^{(\mathrm{log})}}
\newcommand{\thalfbass}{t_{1/2}^{(\mathrm{bass})}}
\newcommand{\thalf}{t_{1/2}}
\newcommand{\npcmodel}{n}
\newcommand{\npc}{\hat n}

\newcommand{\D}{\mathop{}\!{\mathrm d}}

\newcommand{\adjRsq}{R^2_\mathrm{adj}}
\newcommand{\RMSD}{\mathrm{RMSD}}

\begin{document}

\title{Exponential Adoption of Battery Electric Cars}

\author{Felix Jung} 
\affiliation{Chair of Network Dynamics, Center for Advancing Electronics Dresden (cfaed) and Institute of Theoretical Physics, TUD Dresden University of Technology, 01062 Dresden, Germany}

\author{Malte Schröder} 
\affiliation{Chair of Network Dynamics, Center for Advancing Electronics Dresden (cfaed) and Institute of Theoretical Physics, TUD Dresden University of Technology, 01062 Dresden, Germany}

\author{Marc Timme} 
\affiliation{Chair of Network Dynamics, Center for Advancing Electronics Dresden (cfaed) and Institute of Theoretical Physics, TUD Dresden University of Technology, 01062 Dresden, Germany}
\affiliation{Lakeside Labs, 9020 Klagenfurt, Austria}

\date{\today}

\begin{abstract}
\noindent The adoption of battery electric vehicles (BEVs) may significantly reduce greenhouse gas emissions caused by road transport. However, there is wide disagreement as to how soon battery electric vehicles will play a major role in overall transportation. Focusing on battery electric passenger cars, we here analyze BEV adoption across $17$ individual countries, Europe, and the World, and consistently find exponential growth trends. Modeling-based estimates of future adoption given past trends suggests system-wide adoption substantially faster than typical economic analyses have proposed so far. For instance, we estimate the majority of passenger cars in Europe to be electric by about 2031. Within regions, the predicted times of mass adoption are largely insensitive to model details. Despite significant differences in current electric fleet sizes across regions, their growth rates consistently indicate fast doubling times of approximately 15 months, hinting at radical economic and infrastructural consequences in the near future.
\end{abstract}


\maketitle

\section{Introduction}

Climate change poses a major challenge to humanity in the near to mid-term future \cite{ipcc2023}. It is caused by anthropogenic emissions of greenhouse gases, predominantly carbon dioxide \cite{karl2003, hohne2011, wei2016}. To combat climate change, global carbon dioxide emissions need to be reduced significantly and rapidly. A major contributor to greenhouse gas emissions in general and to those of western countries in particular is the road transport sector \cite{chapman2007}, to a large degree due to passenger cars propelled by internal combustion engines \cite{sacchi2022, wuebbles2001}. In 2019, transportation caused $30\,\%$ of the greenhouse gas emissions in the United States of America (USA), and $18\,\%$ in Europe, but only $9\,\%$ in India, and $7\,\%$ in China \cite{ritchie2020}. Over the past decades, however, the transport sector has failed to reduce its carbon footprint \cite{solaymani2019}. 

Mitigating carbon emissions from the transport sector requires the transition to carbon-neutral powertrains, and thus the replacement of mineral oil as the underlying source of energy. 
Today, the two most common options to achieve this goal are to either produce hydrogen by processing biomass or splitting water \cite{nikolaidis2017, dasilvaveras2017}, which is subsequently used to power an electric motor by means of a fuel cell \cite{wilberforce2017}, or generate electric energy from renewable sources to directly power an electric motor by means of a rechargeable (often lithium-ion-based) battery \cite{miao2019}. 

Several related technologies are also under development. For instance, hydrogen \cite{boretti2020} and synthetic fuels, such as \emph{biofuels} or \textit{electrofuels} \cite{ababneh2022, brynolf2018, tatin2016, lester2020}, may be used to power internal combustion engines. Moreover, batteries based on other elements than lithium are under active development, most prominently sodium-ion batteries \cite{vaalma2018, fichtner2022}.

The technologies required to refuel vehicles based on electric batteries, compressed hydrogen, cryogenic liquid hydrogen, and synthetic fuels have little technological overlap \cite{korberg2023, kalghatgi2018} such that several support infrastructure systems would need to be established for multiple powertrain options to coexist. Transitioning from the currently largely homogeneous infrastructure supporting vehicles with petrol- and diesel-based internal combustion engines to a heterogeneous infrastructure would likely incur substantially higher operational costs. Such economic reasons may act as additional drivers to single out just one winning technology that dominates the market. With battery-electric powertrains currently being the most common carbon-neutral option by far, this provokes the question whether they will succeed as the dominant technology, and, if so, when this will happen.

We strive to answer this question by performing a purely data-driven analysis of the recent worldwide adoption of battery electric cars (BECs), i.e. battery electric light passenger vehicles, and put its results into perspective with regard to the total number of registered passenger cars (PCs). We primarily focus our analysis on the major growing BEC markets of Europe and the USA, considering the substantial uncertainty of the future total PC fleet size in developing countries such as China and India. Bearing this uncertainty in mind, however, we still provide aggregate worldwide results.

We demonstrate that the observed recent adoption of battery electric cars has happened exponentially. Based on this result, we describe the historic growth trends using three related models: An exponential model, a logistic model and a Bass diffusion model. We then employ these models to estimate a future transition point in the mobility landscape: The point in time at which battery electric vehicles are expected to start dominating the passenger car fleet in a country, i.e., to comprise $50\,\%$ of the total fleet, \emph{should} the current trend continue.

\section{Data}

Our analysis relies on worldwide historic BEC and PC registration data, described in detail below. The data cover the years from 2011 to 2022, taking the beginning of the year as the reference.

\subsection{Electric Vehicle Stock}

Our data for the number of BECs across various regions have been provided by the International Energy Agency (IEA), an intergovernmental organization, accompanying their \emph{Global EV Outlook 2022} \cite{iea2022a}.
This dataset provides information on the adoption of electric vehicles and public charging infrastructure in 34 regions worldwide. This includes data from 30 countries (Australia, Belgium, Brazil, Canada, Chile, China, Denmark, Finland, France, Germany, Greece, Iceland, India, Indonesia, Italy, Japan, Korea, Mexico, Netherlands, New Zealand, Norway, Poland, Portugal, South Africa, Spain, Sweden, Switzerland, Thailand, USA, United Kingdom) and four aggregate regions (Europe, Other Europe, Rest of the world, World). The dataset contains both year-resolved historical data, ranging from 2011 to 2022, and scenario-based projected future data up to 2031, according to their \emph{Announced Pledges Scenario} (APS) and \emph{Stated Policies Scenario} (STEPS). For several kinds of vehicles (buses, cars, trucks, vans), and the respective powertrain technology (battery electric, plug-in hybrid), sales and stock numbers are provided. The dataset also contains information on charging infrastructure, namely the number of public slow and fast chargers. Finally, for select large regions, values for mineral oil displacement and electricity demand of electric vehicles is provided (China, Europe, India, Rest of the world, USA, World). Not all described data are available for all the countries contained in the dataset. 

For our analysis, we extract the year-resolved historic data on the absolute and relative number of battery electric passenger cars per region.

\subsection{Passenger Car Stock}

To model the total passenger car fleet, we gather vehicle registration data from several sources.

The International Organization of Motor Vehicle Manufacturers (OICA) \cite{oica} provides worldwide vehicle registration data, distinguishing between passenger cars, defined as motor vehicles intended for passenger transport of a maximum capacity of nine persons, and commercial vehicles, such as trucks, coaches, and buses. The dataset contains vehicle stock numbers for the years 2015 and 2020 and 63 countries. For Europe, this includes the categories "EU 27 + EFTA + UK" (Austria, Belgium, Bulgaria, Croatia, Czechia, Denmark, Finland, France, Germany, Greece, Hungary, Ireland, Italy, Netherlands, Norway, Poland, Portugal, Romania, Slovakia, Spain, Sweden, Switzerland, the United Kingdom, and a residual category) and "Russia, Turkey and Other Europe" (Belarus, Russia, Serbia, Turkey, Ukraine, and a residual category), for America, the categories "NAFTA" (Canada, Mexico, USA) and "Central and South America" (Argentina, Brazil, Chile, Colombia, Ecuador, Peru, Venezuela, and a residual category), for Asia, Oceania, and the Middle East, the countries of Australia, China, India, Indonesia, Iran, Iraq, Israel, Japan, Kazakhstan, Malaysia, New Zealand, Pakistan, the Philippines, South Korea, Syria, Taiwan, Thailand, the United Arab Emirates, Vietnam and a residual category, and for Africa, the countries of Algeria, Egypt, Libya, Morocco, Nigeria, South Africa, and a residual category.

For Europe, Eurostat \cite{eurostat2023}, a Directorate-General of the European Commission, provides vehicle registration data. This includes data for countries which are members of the European Union (EU) (Austria, Belgium, Bulgaria, Croatia, Cyprus, Czechia, Denmark, Estonia, Finland, France, Germany, Greece, Hungary, Ireland, Italy, Latvia, Lithuania, Luxembourg, Malta, Netherlands, Poland, Portugal, Romania, Slovakia, Slovenia, Spain, Sweden), Liechtenstein, Norway, and Switzerland, which are members of the European Free Trade Association, North Macedonia, Turkey, which are EU membership candidate countries, Kosovo, which is a potential EU membership candidate country, and the United Kingdom, which is a former EU member. 

The dataset contains passenger car stock numbers for the respective countries for each year from 2012 to 2019, except for a few missing values. The stock numbers are individually specified for each of the surveyed powertrain technologies: bi-fuel, biodiesel, bioethanol, diesel, diesel (excluding hybrids), electricity, hybrid diesel-electric, plug-in hybrid diesel-electric, hybrid electric-petrol, plug-in hybrid petrol-electric, natural gas, hydrogen and fuel cells, liquefied petroleum gases (LPG), petroleum products, petrol (excluding hybrids), and a residual category. These categories are defined in more detail in \cite{europeanunion2019}.

For the USA, the Federal Highway Administration (FHWA) releases annual vehicle registrations data, broken down by state \cite{fhwa2022}. Registrations are reported individually for passenger cars, buses, trucks, and motorcycles. Additionally, the dataset separates private and commercial registrations on the one hand, and public registrations on the other. The most recent data have been released for 2021. We have extracted data in a common format starting from 2007.

For Iceland, Statistics Iceland, the National Statistical Institute of Iceland, releases data on the number of registered passenger cars \cite{statisticsiceland2022}, from 1950 to most recently 2021 (number at end of year).

\subsection{Rapid BEC adoption worldwide}

\begin{figure}
    \includegraphics{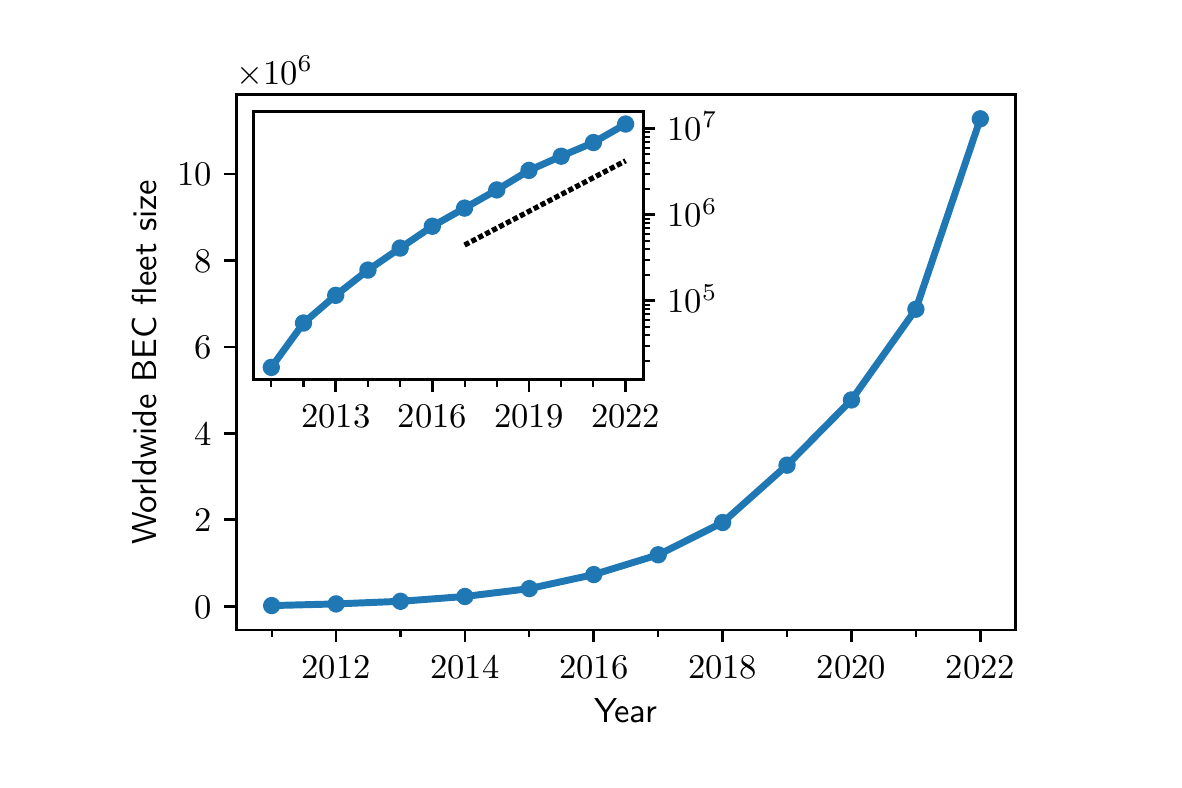}
    \caption{
        \textbf{Global exponential growth of BEC adoption.} 
        The number of battery electric cars in operation has grown rapidly across the world. The straight-line increase on the logarithmic-linear scale (inset) over two orders of magnitude (black line indicates slope of fit of 2016-2022 data) highlights that the recent adoption has happened exponentially.
    }
    \label{fig:worldwide-exponential-adoption}
\end{figure}

Globally, the number of BECs has been growing rapidly, resembling an exponential adoption process, see Figure \ref{fig:worldwide-exponential-adoption}, and reaching a value of approximately 11 million vehicles in the beginning of 2022. 

\begin{figure}
    \includegraphics{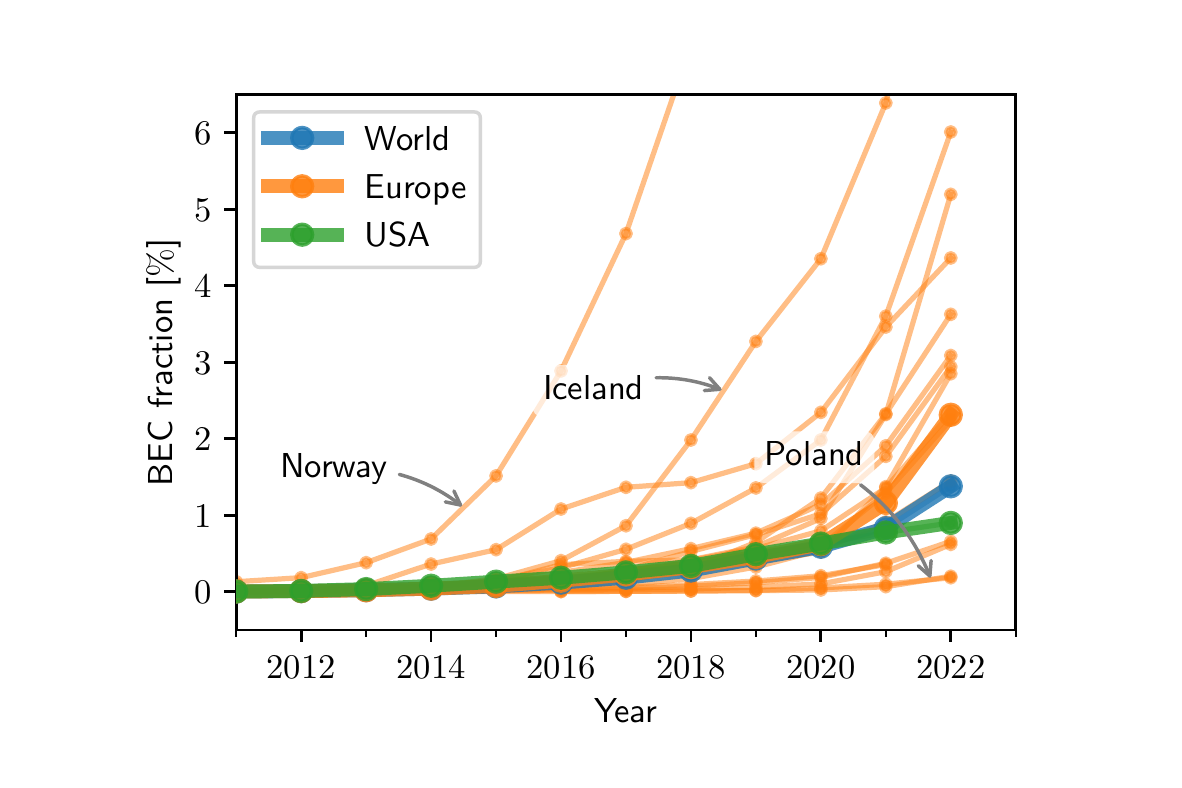}
    \caption{
        \textbf{BEC adoption growth across regions.}
        The BEC share in the total PC fleet has grown significantly over the past decade, with individual countries being at very different stages of adoption: In 2022, Norway features a BEC share of $25.3\,\%$ (out of frame), while Poland is at just under $0.2\,\%$. Europe outperforms both the USA and the world average.
    }
    \label{fig:historic-bec-share-comparison}
\end{figure}

The high absolute numbers are reflected in the BEC fraction of the total PC fleet, see Figure \ref{fig:historic-bec-share-comparison}, reaching approximately $1.4\,\%$ globally in the beginning of 2022, with Europe having a $2.3\,\%$ BEC fraction and the USA $0.9\,\%$. Comparing the adoption in individual countries reveals significant variation in the current BEC fraction values. In Europe, Norway is leading the way at $25.3\,\%$ in the beginning of 2022, while at the same time, Poland has the lowest fraction of just under $0.2\,\%$, spanning a range of two orders of magnitude.

Whether these very different adoption processes follow a common law and what this means for the dominance of BECs in those countries is the subject of our analysis.

\section{Methods}

We model the historic adoption trends of battery electric vehicles in 16 individual European countries (Belgium, Denmark, Finland, France, Germany, Greece, Iceland, Italy, Netherlands, Norway, Poland, Portugal, Spain, Sweden, Switzerland, United Kingdom) and the residual category "Other Europe", containing 17 countries (Austria, Bulgaria, Croatia, Cyprus, Czechia, Estonia, Hungary, Ireland, Latvia, Liechtenstein, Lithuania, Luxembourg, Malta, Romania, Slovakia, Slovenia, Turkey), in the USA, and globally aggregated. 

For each of these regions, country or aggregate, we fit three related growth models, an exponential model $\nexp$, a logistic model $\nlog$, and a bass diffusion model $\nbass$, to the empiric number of registered BECs $\hat n_\mathrm E(t)$ at times $t$ measured in calendar years, ranging from 2011 to 2022. All three models are described in more detail below.

To estimate the point in time at which BECs start dominating the passenger car fleet, we extrapolate the models into the future. We describe the total passenger fleet size using a trivial model $\npcmodel(t)=\npc(\tilde t)=n$, assuming a constant total number of passenger cars since the year 2020, i.e., we take $\tilde t=2020$. Based on data availability, we gather the total PC fleet size from multiple sources: For the USA we use the number reported by the FHWA \cite{fhwa2022}, and for Iceland we use the number reported by Statistics Iceland \cite{statisticsiceland2022}. We use the numbers reported by the International Organization of Motor Vehicle Manufacturers (OICA) for all other countries contained in the dataset. For the remaining countries (Cyprus, Estonia, Iceland, Kosovo, Latvia, Liechtenstein, Lithuania, Luxembourg, Malta, North Macedonia, and Slovenia) we resort to the latest number reported by Eurostat \cite{eurostat2023} for 2019 (for these countries, $\tilde t=2019$).

For the aggregate region "Other Europe", we sum the values for the PC fleet size from the sources described above (summing over Cyprus, Estonia, Iceland, Kosovo, Latvia, Liechtenstein, Lithuania, Luxembourg, Malta, North Macedonia, and Slovenia). 

Similarly, for the aggregate region "Europe", we sum over the individual countries (Belgium, Denmark, Finland, France, Germany, Greece, Iceland, Italy, Netherlands, Norway, Poland, Portugal, Spain, Sweden, Switzerland, United Kingdom) and "Other Europe".

For the aggregate region "World", we directly use the aggregate global value for the year 2020 provided by the OICA.

\begin{figure}
    \includegraphics{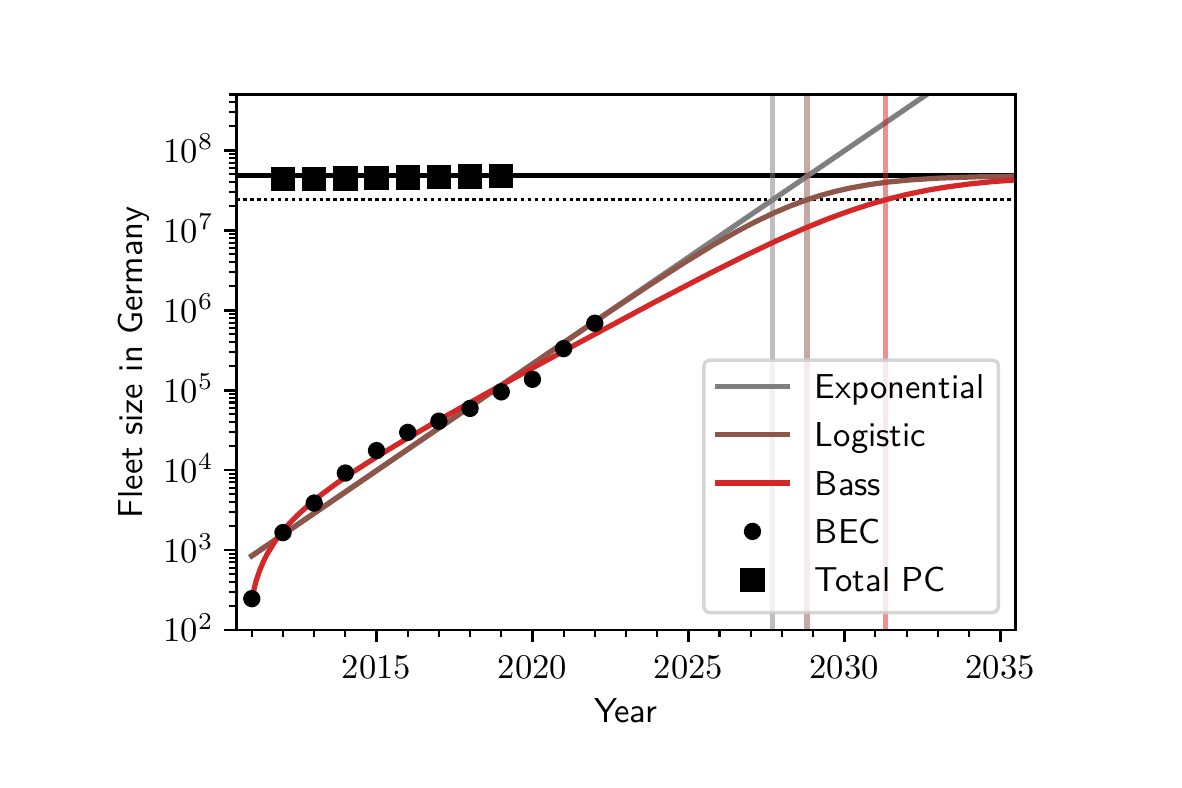}
    \caption{
        \textbf{Estimating BEC dominance.}
        Battery electric car fleet sizes may soon dominate the total passenger car fleet, shown here for Germany as an example. The past total passenger car fleet size stayed approximately constant (black horizontal line), square-black data points taken from the Eurostat dataset. The exponential (gray, Eq.~\eqref{eq:expmodel}), logistic (brown, Eq.~\eqref{eq:logisticmodel}) and Bass diffusion (red, Eq.~\eqref{eq:bassmodel}) models predict the future growth of BEC adoption. The dominance times $\thalf$ (gray, brown, red vertical lines) at which half of the passenger car fleet (gray horizontal line) consists of BECs are estimated to occur between 2027 and 2032, depending on the model. 
    }
    \label{fig:bec-dominance-estimation-germany}
\end{figure}

Using the models for the fleet sizes of BECs $\ngeneral(t)$ and PCs $\npcmodel(t)=n$, we determine the points in time $\thalf$ at which BECs are expected to start dominating the PC fleet, i.e. $n_\mathrm E(t)=n/2$, according to the three fitted models, using the construction shown in Figure \ref{fig:bec-dominance-estimation-germany} for the example of Germany. Here, the historic PC fleet sizes are taken from the Eurostat dataset.

\subsection{Exponential model}

A growth process of a quantity $\nexp(t)$, whose growth rate (i.e., the change of $\nexp$ over a time $t$) is proportional to the quantity itself, is called \emph{exponential}. In its simplest form, this dynamics is formalized by the differential equation
\begin{equation}
    \frac{\D \nexp(t)}{\D t} = a \, \nexp(t)
    \label{eq:ODEexponential}
\end{equation}
with growth rate $a>0$.

Its solution motivates our \emph{exponential model} for the number of BECs,
\begin{equation}
    \nexp(t) = \exp{\left(a(t-t_0)\right)}\qquad a,t_0,t>0,
    \label{eq:expmodel}
\end{equation}
with a scaled temporal offset $t_0$.

Assuming multiplicative measurement errors, we perform a non-linear least squares regression of the logarithmized version
\begin{equation}
    \log\nexp(t) = a(t-t_0)
    \label{eq:logexpmodel}
\end{equation}
of the model to estimate the parameters $a$ and $t_0$.

\subsection{Logistic model}

While the exponential model described above is well suited for modeling initial adoption, it lacks the idea of ultimate saturation, mathematically suggesting indefinite growth. The logistic model introduces such a saturation limit by making the growth rate of a quantity $\nlog$ additionally proportional to the difference between the quantity and that limit. In its simplest form, the process is characterized by modifying the differential equation \eqref{eq:ODEexponential} to become
\begin{equation}
    \frac{\D \nlog}{\D t} = a \, \nlog \left(1-\nlog\right).
    \label{ODElogistic}
\end{equation}

Its solution motivates our \emph{logistic model} for the number of BECs, given by
\begin{equation}
    \nlog(t) = \frac{L}{1+\exp\left(-a(t-t_0) \right)}\qquad L, a,t_0,t>0,
    \label{eq:logisticmodel}
\end{equation}
with an initial growth rate $a$, a time offset $t_0$ and the saturation limit $L$.

The logistic model is an established tool to describe technology adoption, having been employed in a number of studies \cite{griliches1957, frank2004, gruber2001, lee2007, liikanen2004}. 

Like for the exponential model, we assume multiplicative measurement errors and perform the parameter optimization of the parameters $a$ and $t_0$ on the logarithmized form
\begin{equation}
    \log\nlog(t) = \log\left(\frac{L}{1+\exp\left(-a(t-t_0) \right)}\right),
    \label{eq:loglogmodel}
\end{equation}
by non-linear least squares regression. We keep the saturation limit $L$ fixed to our constant model of the total PC fleet size, i.e. $L=n$.

\subsection{Bass diffusion model}

Explicitly modeling consumer behavior during the adoption of new products, Bass \cite{bass1969} derived the \emph{Bass diffusion model}, treating both initial and replacement purchases, and classifying customers as \emph{innovators} or \emph{imitators}. The Bass model is widely used to describe innovation diffusion \cite{qian2014, sundqvist2005, wu2010, bottomley1998, dekimpe1998}.

It describes the growth of the number of BECs $\nbass$ by the differential equation
\begin{equation}
\frac{\D \nbass(t)}{\D t} = p(L-\nbass(t))+\frac{q}{L}\nbass(t)(L-\nbass(t))\qquad L,q,p>0,
\end{equation}
with the \textit{coefficient of innovation} $p$, the \textit{coefficient of imitation} $q$, and the saturation limit $L$. The first term, $p(L-\nbass(t))$, describes adoption due to the \textit{innovators}, the second term, $\frac{q}{L}\nbass(t)(L-\nbass(t))$ represents adoption due to the \textit{imitators} \cite{jukic2011}.

Solving the equation with $\nbass(0)=0$, we use it to model the number of BECs in the following form:

\begin{equation}
    \nbass(t) = L\frac{1-\exp\left(-(p+q)(t-t_0)\right)}{1+qp^{-1}\exp\left(-(p+q)(t-t_0)\right)}\qquad
    L,q,p,t_0,t>0,
    \label{eq:bassmodel}
\end{equation}
with adoption parameters $p$ and $q$, a temporal offset $t_0$ and the saturation limit $L$. 

As before, we assume multiplicative measurement errors and perform the parameter optimization of the parameters $p$, $q$ and $t_0$ on the logarithmized form
\begin{equation}
    \log\nbass(t) =
    \log\left(L\frac{1-\exp(-(p+q)(t-t_0))}{1+qp^{-1}\exp(-(p+q)(t-t_0))}\right),
    \label{eq:logbassmodel}
\end{equation}
using non-linear least squares regression. Like for the logistic model, we keep the saturation limit $L$ fixed to our constant model of the total PC fleet size, i.e. $L=n$.

\section{Results}

We report two types of results, one on extracting past trends of BEC adoption, as provided by fitting the three classes of models to past data, and the second on estimating future BEC adoption, should the past trends continue according to these models. The numerical results of our model fitting procedure are listed in Table \ref{tab:params}, which for all countries provides the optimal parameter estimates, as well as the respective values of $\adjRsq$ and the root-mean-square deviation ($\RMSD$).

\begin{turnpage}
\begin{table}
    \begin{tabular}{l|l|l|l|l|l|l|l|l|l|l|l|l|l|l|l|}
\toprule
Model & \multicolumn{4}{|c|}{Exponential} & \multicolumn{5}{|c|}{Logistic} & \multicolumn{6}{|c|}{Bass} \\
Parameter type & \multicolumn{2}{|c|}{Fit} & \multicolumn{2}{|c|}{Properties} & \multicolumn{2}{|c|}{Fit} & \multicolumn{1}{|c|}{Fixed} & \multicolumn{2}{|c|}{Properties} & \multicolumn{3}{|c|}{Fit} & \multicolumn{1}{|c|}{Fixed} & \multicolumn{2}{|c|}{Properties} \\
Parameter & $a$ & $t_0$ & $R^2_\mathrm{adj}$ & $\mathrm{RMSD}$ & $a$ & $t_0$ & $L$ & $R^2_\mathrm{adj}$ & $\mathrm{RMSD}$ & $p$ & $q$ & $t_0$ & $L$ & $R^2_\mathrm{adj}$ & $\mathrm{RMSD}$ \\
\midrule
World & $0.56$ & $1992.30$ & $0.79$ & $1372054$ & $0.56$ & $2029.74$ & $1180677016$ & $0.80$ & $1356589$ & $2.16\times 10^{-5}\,$ & $0.47$ & $2010.44$ & $1180677016$ & $0.99$ & $319369$ \\
Europe & $0.51$ & $1992.69$ & $0.98$ & $121609$ & $0.51$ & $2030.84$ & $307157761$ & $0.98$ & $119542$ & $2.20\times 10^{-5}\,$ & $0.44$ & $2010.16$ & $307157761$ & $0.98$ & $107877$ \\
Belgium & $0.53$ & $2001.29$ & $0.94$ & $3463$ & $0.53$ & $2030.47$ & $5827195$ & $0.94$ & $3413$ & $3.06\times 10^{-5}\,$ & $0.41$ & $2010.67$ & $5827195$ & $0.97$ & $2375$ \\
Denmark & $0.51$ & $2000.09$ & $0.92$ & $4557$ & $0.51$ & $2029.15$ & $2720273$ & $0.92$ & $4560$ & $1.23\times 10^{-4}\,$ & $0.34$ & $2010.83$ & $2720273$ & $0.80$ & $6993$ \\
Finland & $0.57$ & $2004.93$ & $0.90$ & $1908$ & $0.57$ & $2030.95$ & $2748448$ & $0.89$ & $1923$ & $1.00\times 10^{-7}\,$ & $0.57$ & $2003.64$ & $2748448$ & $0.88$ & $1931$ \\
France & $0.55$ & $1997.35$ & $0.54$ & $80141$ & $0.55$ & $2029.32$ & $38458212$ & $0.55$ & $79591$ & $7.02\times 10^{-5}\,$ & $0.37$ & $2010.89$ & $38458212$ & $0.98$ & $14425$ \\
Germany & $0.61$ & $2000.02$ & $0.97$ & $31237$ & $0.62$ & $2028.81$ & $48248584$ & $0.97$ & $30958$ & $2.16\times 10^{-5}\,$ & $0.49$ & $2010.78$ & $48248584$ & $0.88$ & $56284$ \\
Greece & $0.70$ & $2010.58$ & $0.95$ & $201$ & $0.70$ & $2032.80$ & $5315875$ & $0.95$ & $201$ & $3.72\times 10^{-6}\,$ & $0.47$ & $2013.85$ & $5315875$ & $0.61$ & $496$ \\
Iceland & $0.64$ & $2006.83$ & $0.52$ & $1930$ & $0.65$ & $2026.24$ & $269825$ & $0.56$ & $1859$ & $2.31\times 10^{-5}\,$ & $0.60$ & $2009.87$ & $269825$ & $0.78$ & $1228$ \\
Italy & $0.45$ & $1997.33$ & $0.75$ & $15569$ & $0.45$ & $2036.02$ & $39717874$ & $0.75$ & $15589$ & $1.00\times 10^{-7}\,$ & $0.45$ & $2002.08$ & $39717874$ & $0.71$ & $15664$ \\
Netherlands & $0.58$ & $2000.19$ & $0.96$ & $14398$ & $0.58$ & $2027.94$ & $9049959$ & $0.96$ & $13561$ & $3.31\times 10^{-5}\,$ & $0.51$ & $2010.21$ & $9049959$ & $0.96$ & $13957$ \\
Norway & $0.48$ & $1993.83$ & $0.77$ & $60735$ & $0.50$ & $2024.58$ & $2794457$ & $0.84$ & $50223$ & $5.07\times 10^{-4}\,$ & $0.43$ & $2009.77$ & $2794457$ & $0.97$ & $22185$ \\
Poland & $0.62$ & $2006.17$ & $1.00$ & $125$ & $0.62$ & $2033.61$ & $25113862$ & $1.00$ & $126$ & $1.25\times 10^{-7}\,$ & $0.62$ & $2008.75$ & $25113862$ & $1.00$ & $135$ \\
Portugal & $0.38$ & $1994.81$ & $0.88$ & $3500$ & $0.38$ & $2035.99$ & $5300000$ & $0.88$ & $3519$ & $1.00\times 10^{-7}\,$ & $0.38$ & $1995.72$ & $5300000$ & $0.86$ & $3525$ \\
Spain & $0.54$ & $2001.01$ & $0.90$ & $6008$ & $0.54$ & $2032.40$ & $25169158$ & $0.90$ & $5983$ & $9.29\times 10^{-6}\,$ & $0.42$ & $2010.69$ & $25169158$ & $0.99$ & $2105$ \\
Sweden & $0.74$ & $2005.55$ & $0.30$ & $23784$ & $0.74$ & $2026.40$ & $4944067$ & $0.32$ & $23364$ & $3.04\times 10^{-5}\,$ & $0.53$ & $2010.97$ & $4944067$ & $0.98$ & $4031$ \\
Switzerland & $0.53$ & $1999.98$ & $0.81$ & $9921$ & $0.53$ & $2028.95$ & $4728444$ & $0.82$ & $9736$ & $5.86\times 10^{-5}\,$ & $0.43$ & $2010.50$ & $4728444$ & $0.99$ & $1606$ \\
United Kingdom & $0.47$ & $1995.13$ & $0.94$ & $24991$ & $0.47$ & $2032.04$ & $36454665$ & $0.94$ & $25324$ & $2.67\times 10^{-6}\,$ & $0.46$ & $2006.24$ & $36454665$ & $0.92$ & $27104$ \\
Other Europe & $0.54$ & $1999.46$ & $0.93$ & $11486$ & $0.54$ & $2032.27$ & $50296863$ & $0.93$ & $11431$ & $1.06\times 10^{-5}\,$ & $0.42$ & $2010.70$ & $50296863$ & $0.97$ & $6950$ \\
USA & $0.50$ & $1992.48$ & $0.18$ & $363801$ & $0.50$ & $2029.39$ & $105135300$ & $0.20$ & $361173$ & $1.02\times 10^{-4}\,$ & $0.36$ & $2010.69$ & $105135300$ & $0.90$ & $118900$ \\
\bottomrule
\end{tabular}

    \caption{
        \textbf{BEC growth model parameters and fit properties.} 
        Model parameters, $\adjRsq$, and root-mean-square deviation $\mathrm{RMSD}$ for the three models considered, exponential, logistic, and Bass diffusion.
    }
    \label{tab:params}
\end{table}
\end{turnpage}

Past trends confirm current exponential BEC adoption (compare Figure \ref{fig:worldwide-exponential-adoption}). The exponential model matches the current BEC adoption in most regions with high accuracy (Table \ref{tab:params}) despite the large differences in the current fleet penetration of BECs across countries (compare Figure \ref{fig:historic-bec-share-comparison}). Importantly, it suggests current growth rates similar to the logistic model that includes a saturation limit. These results highlight that current growth is close to purely exponential and saturation effects are still negligible. 

\begin{figure}
    \includegraphics{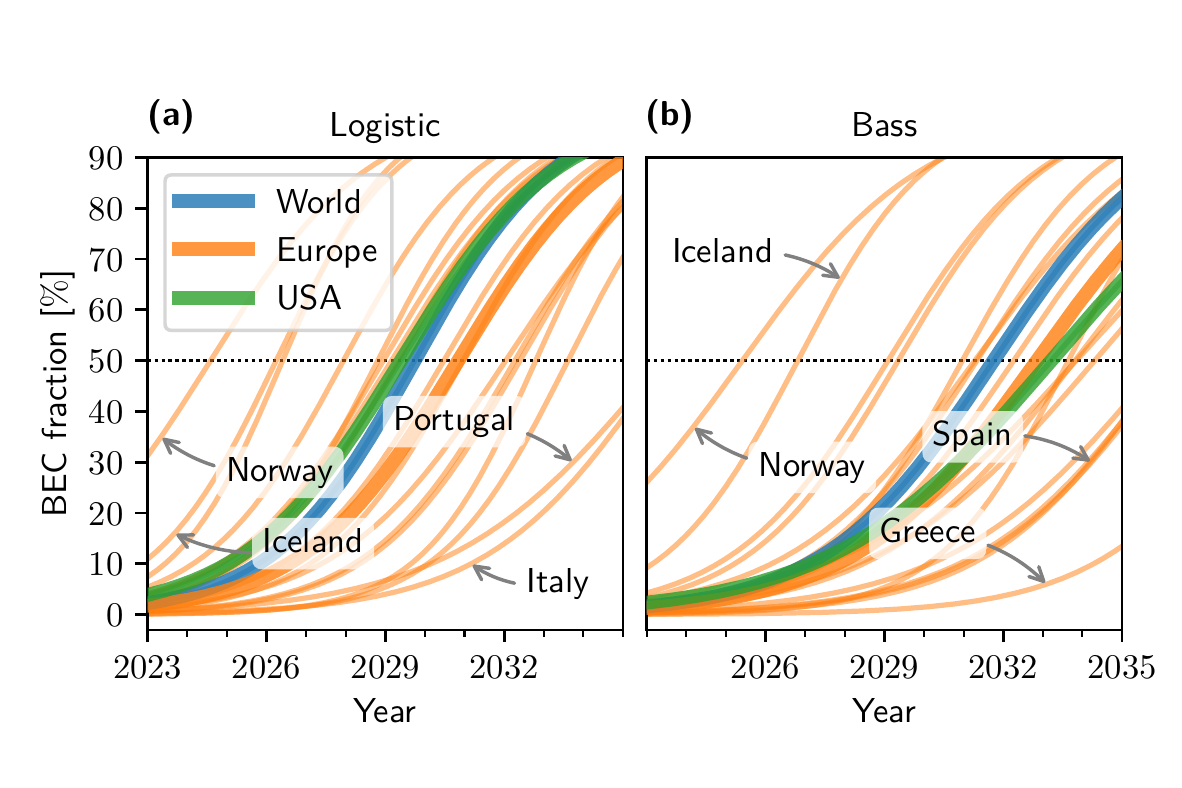} 
    \caption{
        \textbf{Model-estimated BEC adoption trajectories.} 
        The speed of BEC adoption varies strongly across countries (thin lines) in both the logistic model \textbf{(a)} and the Bass model \textbf{(b)}. Norway, Sweden, and the Netherlands exhibit the fastest adoption, Spain, Portugal, Italy, and Greece the slowest.
        Overall, the models predict rapid dominance of BECs (more than $50\,\%$ of the PC fleet, horizontal gray line) in Europe, the USA and globally shortly after 2030.
    }
    \label{fig:bec-adoption-curves}
\end{figure}

When predicting the possible adoption pathways farther into the future, saturation effects play a more significant role. 
Figure \ref{fig:bec-adoption-curves} shows the resulting adoption trajectories estimated by the logistic and Bass models. The Bass model consistently provides more conservative trajectories than the logistic model. Within the models' respective families of adoption curves, the large aggregate regions (World, Europe, USA) exhibit similar adoption dynamics, while the individual European countries differ significantly with respect to the current stage of BEC adoption. While some countries, such as Norway and Iceland, are estimated to undergo a transition to BEC dominance in the near future, others, such as Italy and Portugal, are estimated to reach the BEC dominance about a decade later. These differences apply similarly for both the logistic and Bass models.

\begin{figure}
    \includegraphics{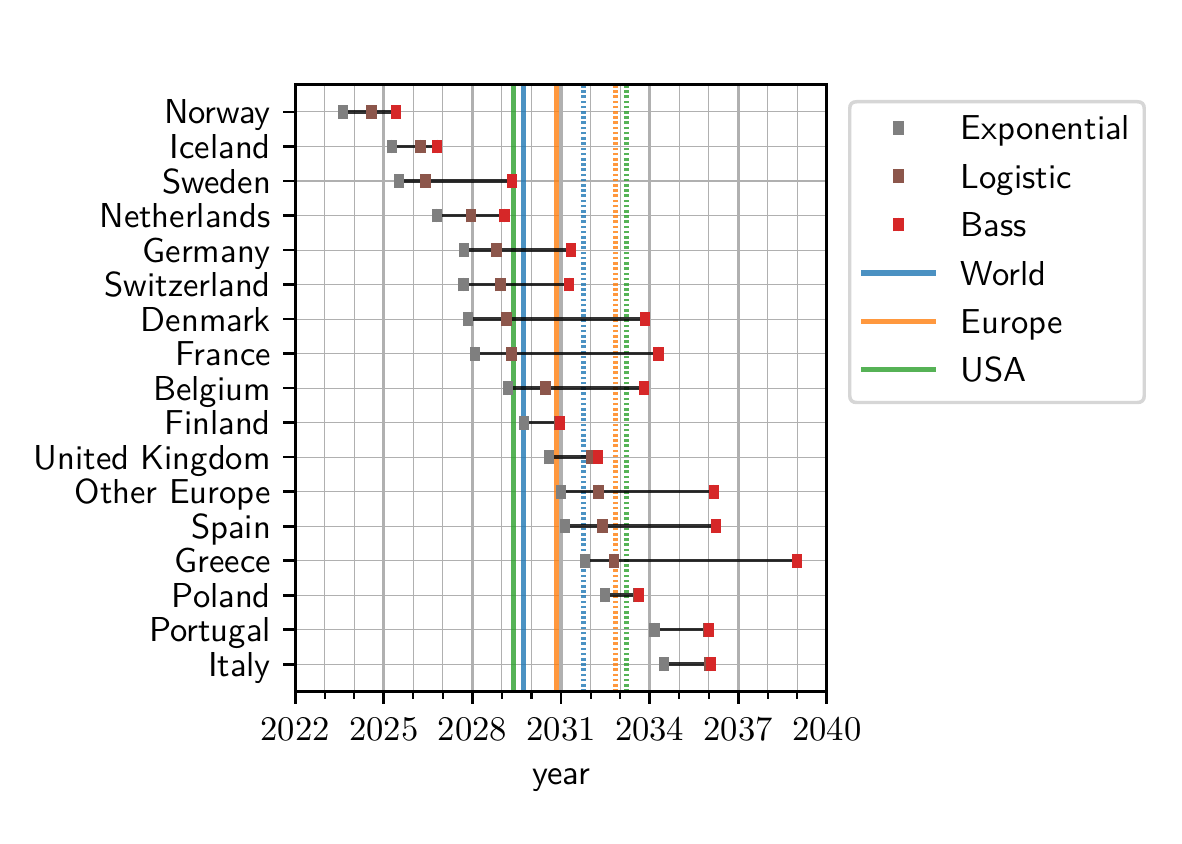}
    \caption{
        \textbf{BEC dominance time estimates.} 
        Battery electric cars start to dominate the passenger car fleet in European countries between 2023 and 2039, according to our models. The ranking of the estimated dominance times of the countries is consistent across the three models, with the exponential model giving the earliest time, the logistic model an intermediate time, and the Bass diffusion model yielding the latest time. The dominance time estimates for the USA and the aggregate regions \textit{Europe} and \textit{World} are indicated by vertical lines, according to the logistic (solid lines) and Bass diffusion (dotted lines) models.
    }
    \label{fig:bec-dominance-estimation-comparison}
\end{figure}

The spread in the stage of individual regions within their transition to BEC dominance is reflected in the estimated dominance times $\thalf$ visualized in Figure \ref{fig:bec-dominance-estimation-comparison}, computed for each of the three BEC models (exponential, logistic, Bass diffusion). Table \ref{tab:domest} lists the corresponding numeric values. For each region, the three models predict the transition to BEC dominance in a consistent rank order, with the exponential models yielding the earliest estimates, the Bass diffusion models the most conservative estimates and the logistic models intermediate ones, $\thalfexp\leq\thalflog\leq\thalfbass$. Despite the lack of a saturation limit in the former, the estimates of the exponential models and the logistic models are consistently close to each other, with differences of the order of one year. This again highlights the approximate validity of assuming exponential growth up to the point of dominance, as a lowest-order estimate.

\begin{figure}
    \includegraphics{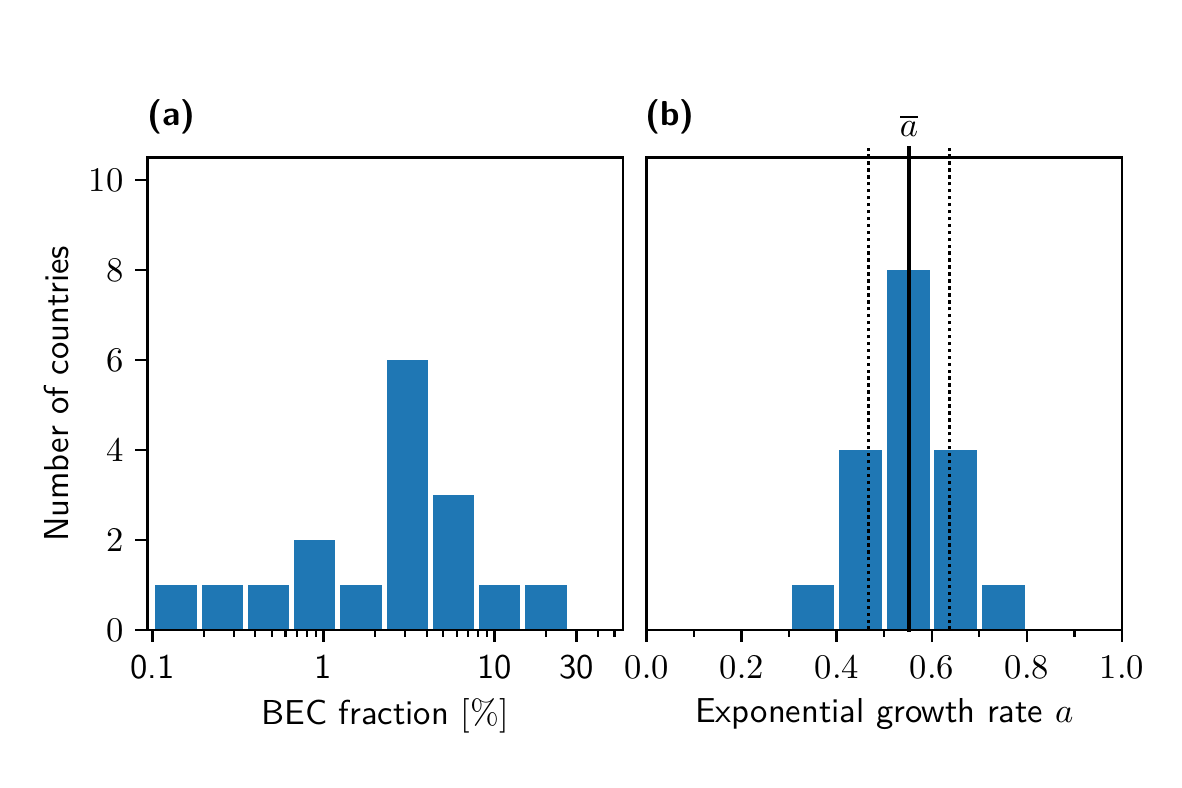}
    \caption{
        \textbf{Stage and growth of BEC adoption across countries.}
        \textbf{(a)} The current (2022) stage of BEC adoption strongly varies across individual countries, ranging from $0.18\,\%$ in Poland to $25.3\,\%$ in Norway (note the logarithmically-scaled abscissa). \textbf{(b)} All countries exhibit similarly large growth rates near the average $\overline{a}=0.55$ per year (solid vertical line), yielding a standard deviation of $\sigma_a=0.09$ per year (dotted vertical lines). 
    }
    \label{fig:bec-fraction-and-growth-rate-distributions}
\end{figure}

In spite of the individual countries representing distinctly different stages of the transition to BEC dominance (Figure \ref{fig:bec-fraction-and-growth-rate-distributions} a), their exponential growth rates $a$ are remarkably similar, with arithmetic mean $\bar a=0.55$ per year and standard deviation of just $\sigma_a=0.09$ per year (Figure \ref{fig:bec-fraction-and-growth-rate-distributions} b). This finding amounts to a doubling time of typically much less than two years ($1.26$ years on average) in the early phase of adoption.

The dominance time estimates of the large aggregate regions (World, Europe, USA) according to the logistic and Bass model (vertical lines in Figure \ref{fig:bec-dominance-estimation-comparison}) are close to each other, spanning a range of only approximately 1.5 years for each model (2029 to 2031 in the logistic, 2031 to 2033 in the Bass model). The dominance time estimate for the World is to be taken with a grain of salt, though, as the assumption of a constant PC fleet size $\npcmodel(t)$ is unlikely to hold, in particular in developing countries with small current passenger car fleet sizes.

\begin{table}
    \begin{tabular}{l|l|l|l|l|l|l|}
\toprule
 & \multicolumn{2}{|c|}{Exponential} & \multicolumn{2}{|c|}{Logistic} & \multicolumn{2}{|c|}{Bass} \\
 & $\thalf^{(\mathrm{exp})}\,[\mathrm{a}]$ & CI-95 $[a]$ & $\thalf^{(\mathrm{log})}\,[\mathrm{a}]$ & CI-95 $[a]$ & $\thalf^{(\mathrm{bass})}\,[\mathrm{a}]$ & CI-95 $[a]$ \\
\midrule
World & $2029$ & $[2027,2030]$ & $2030$ & $[2028,2031]$ & $2032$ & $[2031,2033]$ \\
Europe & $2030$ & $[2028,2031]$ & $2031$ & $[2029,2032]$ & $2033$ & $[2031,2034]$ \\
Belgium & $2029$ & $[2027,2032]$ & $2030$ & $[2028,2033]$ & $2034$ & $[2032,2035]$ \\
Denmark & $2028$ & $[2025,2031]$ & $2029$ & $[2026,2032]$ & $2034$ & $[2030,2037]$ \\
Finland & $2030$ & $[2029,2031]$ & $2031$ & $[2030,2032]$ & $2031$ & -- \\
France & $2028$ & $[2025,2032]$ & $2029$ & $[2026,2032]$ & $2034$ & $[2032,2036]$ \\
Germany & $2028$ & $[2026,2030]$ & $2029$ & $[2026,2031]$ & $2031$ & $[2029,2033]$ \\
Greece & $2032$ & $[2029,2037]$ & $2033$ & $[2029,2036]$ & $2039$ & $[2033,2043]$ \\
Iceland & $2025$ & $[2024,2027]$ & $2026$ & $[2024,2028]$ & $2027$ & $[2025,2029]$ \\
Italy & $2034$ & $[2032,2037]$ & $2036$ & $[2033,2038]$ & $2036$ & -- \\
Netherlands & $2027$ & $[2025,2028]$ & $2028$ & $[2026,2029]$ & $2029$ & $[2027,2031]$ \\
Norway & $2024$ & $[2022,2025]$ & $2025$ & $[2023,2026]$ & $2025$ & $[2024,2027]$ \\
Poland & $2032$ & $[2032,2033]$ & $2034$ & $[2033,2035]$ & $2034$ & -- \\
Portugal & $2034$ & $[2031,2038]$ & $2036$ & $[2032,2039]$ & $2036$ & -- \\
Spain & $2031$ & $[2029,2034]$ & $2032$ & $[2030,2035]$ & $2036$ & $[2035,2038]$ \\
Sweden & $2026$ & $[2023,2029]$ & $2026$ & $[2024,2029]$ & $2029$ & $[2028,2030]$ \\
Switzerland & $2028$ & $[2026,2030]$ & $2029$ & $[2027,2031]$ & $2031$ & $[2029,2033]$ \\
United Kingdom & $2031$ & $[2029,2032]$ & $2032$ & $[2031,2033]$ & $2032$ & -- \\
Other Europe & $2031$ & $[2029,2034]$ & $2032$ & $[2030,2034]$ & $2036$ & $[2035,2038]$ \\
USA & $2028$ & $[2026,2031]$ & $2029$ & $[2027,2032]$ & $2033$ & $[2031,2036]$ \\
\bottomrule
\end{tabular}

    \caption{
        \textbf{BEC dominance time estimates.} 
        Times $\thalf$ in years at which the three models considered, exponential
        (\textit{exp}), logistic (\textit{log}), and Bass diffusion (\textit{bass}),
        predict half of the passenger-car fleet in European countries to be battery electric, and the respective rounded $95\,\%$-confidence intervals. For some of the regions ("--"), the Bass model regression procedure returned no finite error estimate due to the small number of data points. 
    }
    \label{tab:domest}
\end{table}

\section{Discussion}

Our analysis demonstrates that battery electric vehicles have been adopted exponentially across the globe for the past decade. 
Despite the differences in current BEC fleet sizes, both the qualitative exponential trends and the growth rates are highly consistent among countries and regions. The results suggest worldwide BEC dominance in less than a decade into the future, potentially much earlier. Exact numerical values of future BEC numbers are impossible to capture in any modeling setting and will depend on several economic, societal, political and technological factors (see below for further discussion). Such basic modeling is specifically neglecting (unpredictable) future incentives or disincentives and technological advances, not just specifically in the realm of battery-electric powertrains but also generally for cars and other modes of transport.

The modeling of future adoption \textit{numbers} reported above may thus be seen as extending past trends and suggesting future numbers by order of magnitude.
The expected \textit{time} of BEC dominance is largely insensitive to model details as long as a strongly superlinear adoption trend prevails. For instance, for exponential growth patterns in fleet size, the time of adoption at a certain BEC fraction, say $50\,\%$, only logarithmically varies with the predicted number of BECs. For the logistic and Bass models, the dependence of dominance times on total adoption numbers are also approximately logarithmic and such that time point estimates within the model classes are similarly insensitive.
 
As stated, while the currently observed absolute fleet sizes and the BEC shares in the total passenger car fleet differ significantly among regions, all of them exhibit exponential growth patterns, at similar exponential growth rates $a$. Country-level analysis suggests the same trends for all countries with substantial numbers of BEC fleet sizes to start with. These results indicate a global transition towards cars with battery-electric powertrains, with individual countries and regions currently being at different stages of this transition.

In countries worldwide, governments are pushing for a phase-out of fossil-fuel-powered vehicles \cite{wappelhorst2021} to achieve greenhouse gas emission reduction targets. 
For the European Union, aiming for carbon neutrality in 2050, the European Parliament and Council have agreed to ensure that "all new cars and vans registered in Europe will be zero-emission by 2035" \cite{europeancommission2022}. This is binding for the 27 EU member states and, pending adoption by the EEA Joint Committee, to some or all EEA European Free Trade Association (EFTA) states \cite{icct2023}.
For the USA, no federal phase-in target has been enacted except for government-owned vehicles, but individual states (California, Massachusetts, New York, Oregon, Vermont, and Washington) have committed to restrict sales of new passenger cars to battery electric, fuel cell electric, and plugin-hybrid electric vehicles \cite{icct2023}.

While these government commitments will partially enforce a transition to carbon-neutral powertrains, several options of homogeneous or heterogeneous technology adoption are still discussed to date. These include electric powertrains using batteries or hydrogen fuel cells as means of energy storage, and internal combustion engines using electrofuels. Additionally, if and how fast these technologies will make up a significant part of the fleet and whether the transition will be driven by consumer behavior or by government restrictions is subject to debate.
That debate is reflected in research and industry forecasts \cite{brown2020, kah2019}, as well as by those of market analysts \cite{rethinkx2017}. 

In contrast to the results of our data-driven modeling reported above, earlier work by Statharas et al.~estimates the total EV share (both battery electric and plugin-hybrid electric vehicles) in the EU at only up to $21\,\%$ by 2030 through simulations employing agent-based transport modeling \cite{statharas2019}. Rietmann et al.~already predict $30\,\%$ of passenger cars to be EVs by 2032, with much higher numbers for several "Fast EV penetration" countries (Austria, Denmark, Portugal, UK, Germany, Netherlands) \cite{rietmann2020}.

Current longer term projections of business intelligence firms and market analysts polarize heavily.
The management consulting firm McKinsey \& Company predicts rapid BEV adoption and market dominance near the turn of the decade \cite{kaas2016}, the strategic market research provider BloombergNEF similarly expects $50\,\%$ of all cars on the road to be electric in Europe and China by 2030 \cite{pvbnef}. The investment bank Morgan Stanley anticipates only $22\,\%$ of the distance traveled by car in the USA in 2035 to be realized by electric vehicles, and $50\,\%$ by 2040, with $25\,\%$ of new car sales being electric vehicles by 2030 \cite{elkins2023}, surprisingly low numbers given the current adoption and our model estimates.

In their Global EV Outlook 2023, the International Energy Agency (IEA) projects the worldwide BEC share for 2030 at $15\,\%$, according to both of their model scenarios, \emph{Stated Policies Scenario} (STEPS), and \emph{Announced Pledges Scenario} (APS) \cite{iea2023a, iea2023}, compared to our result of $31\,\%$, following the Bass model, which resembles the slowest adoption within our models. For Europe, the IEA predicts $18\,\%$, compared to our $22\,\%$, and for the USA, the IEA predicts $16\,\%$, compared to our $24\,\%$.

Also in contrast to our findings, business intelligence provider Visiongain forecasts that "1 in 12 vehicles sold in California, Germany, South Korea, \& Japan, should be hydrogen-powered" \cite{visiongain2022}, and the business analytics platform MarketWatch states their opinion that "Electric vehicles aren't going to take over any time soon" \cite{brandus2021}.

Car manufacturers' positions on BECs differ widely. In 2019, Honda CEO Takahiro Hachigo still claimed that "Electric vehicles will not go mainstream" and stated that Honda will focus on gasoline-electric hybrids, not on battery electric vehicles, through 2030 \cite{greimel2019}. Similarly, Toyota Motor Corporation's former CEO Akio Toyoda has criticized an "excessive hype over electric vehicles" \cite{landers2020}. Toyota will not focus on BECs, but believe that "non-electric cars will also play a lasting role in global auto markets" for the next 30 years, suggesting that "different options including hybrids and fuel-cell vehicles" are needed "to compete against each other" \cite{davis2021}. Contrary to that, a multitude of other carmakers fully embrace BEC technology, with Audi expecting one third of their annual sales to be fully electric by 2025, and Volkswagen aiming for $70\,\%$ by 2030 \cite{poliscanova2021, adac_bev}.

Considering the current state of carbon-neutral powertrain adoption, however, it becomes clear that so far, BEC adoption has by far outperformed all other technologies. For example in Europe, according to Eurostat data, in 2019, BECs comprised $0.34\,\%$ of the total passenger car fleet. The second and third-largest fractions of cars with carbon-neutral powertrains, powered by bioethanol and by hydrogen fuel cells, comprised only $0.02\,\%$ and $0.0003\,\%$, respectively.

Thus, with battery electric cars being the only potentially carbon-neutral powertrain option that is currently deployed at large scale, it stands to reason that in the upcoming years, hydrogen fuel cells or the aforementioned related technologies will not be able to catch up and hinder BEC adoption, see also \cite{albatayneh2023}. For instance, even at an estimated $69\,\%$ compound annual growth rate (CAGR) \cite{marketresearchfuture2023}, the BEC fleet in Europe in 2029 will outnumber the fleet of vehicles powered by hydrogen fuel cells by more than three orders of magnitude, based on the data that Eurostat reported for 2019.

In our opinion, even if it is not impossible for hydrogen fuel cells or future technologies to play a significant role eventually---the fact that today those technologies are either niche or non-existent gives the exponentially growing BEC market a significant head start, limiting the influence of competing technologies in the mid-term future. Although this argument may not directly apply to regions where the current BEC share is still very small, chances seem high that due to the globalized market, reduction of production costs and sales prices due to mass adoption in other regions will also affect those regions that are still at early stages of BEC adoption, thereby making BECs the most attractive option for customers and manufacturers.

In summary, our data-driven analysis clearly demonstrates that the current initial phase of worldwide electric vehicle adoption reflects genuine exponential growth. Such fast nonlinear adoption dynamics can be attributed to or even further accelerated by cascades of tipping points, triggered through reinforcing feedbacks, such as policy support, learning-by-doing, economies of scale and the emergence of complementary technologies such as charging infrastructure \cite{sharpe2021}.

Via such diverse factors, the characteristic timescale $a^{-1}$ determining the growth rate may change in the future. For instance, slow implementation of charging infrastructure or surging development of alternative emission-free drivetrains, based e.g., on hydrogen, may in principle slow the adoption. In contrast, the current increasing policy support in favor of electric cars, a simpler implementation of autonomous driving for electric vehicles and further commitments of manufacturers or countries to phase out internal combustion engines may accelerate BEC adoption even more.

We conclude that not only are battery electric cars currently adopted exponentially, they are highly likely to dominate the global passenger car fleet in the near future.

\appendix

\section{Funding}
This research was supported by the Bundesministerium für Bildung und Forschung (BMBF, Federal Ministry of Education and Research) under grant No. 16ICR01.

\section{Availability of data and materials}
The data analyzed in this paper are freely available from the respective providers: IEA \cite{iea2022a, iea2023, iea2023a}, OICA \cite{oica}, Eurostat \cite{eurostat2023, europeanunion2019}, FHWA \cite{fhwa2022}, and Statistics Iceland \cite{statisticsiceland2022}.

\section{Competing interests}
The authors declare no competing interests.

\section{Authors' contributions}
M.T. conceived the research, F.J., M.S. and M.T. designed the research. F.J. aggregated and analyzed the data, evaluated the models and prepared all figures, supervised by M.S. and M.T. All authors discussed and interpreted the results and wrote the manuscript.


%

\end{document}